\documentclass[10pt]{article}
\usepackage{amssymb}

\begin{document}

\title{Infinite Hopf Families of Algebras and \\
Yang-Baxter Relations}
\author{Niall MacKay${}^{1}$\thanks{%
e-mail: nm15@york.ac.uk} \hspace{0.5cm} and \hspace{0.5cm} Liu Zhao$%
{}^{1,2,3}$\thanks{%
e-mail: lzhao@phy.nwu.edu.cn} {}\thanks{%
Royal Society visiting fellow} \\
${}^1$ Department of Mathematics, University of York, York YO10 5DD, UK\\
${}^2$ Institute of Modern Physics, Northwest University, Xian 710069, China%
\\
${}^3$ The Abdus-Salam International Centre for Theoretical Physics, 34100
Trieste, Italy}
\date{}
\maketitle

\begin{abstract}
A Yang-Baxter relation-based formalism for generalized quantum
affine algebras with the structure of an infinite Hopf family of
(super-) algebras is proposed. The structure of the infinite Hopf
family is given explicitly on the level of $L$ matrices. The
relation with the Drinfeld current realization is established in
the case of $4\times4$ $R$-matrices by studying the analogue of
the Ding-Frenkel theorem. By use of the concept of algebra
``comorphisms'' (which generalize the notion of algebra comodules
for standard Hopf algebras), a possible way of constructing
infinitely many commuting operators out of the generalized $RLL$
algebras is given. Finally some examples of the generalized $RLL$
algebras are briefly discussed.
\end{abstract}

\section{Introduction}

Yang-Baxter relations ($RTT$ relations) have proved to be a
central ingredient in 2D integrable models and quantum algebras
over the last 20 years. In quantum algebras, the Yang-Baxter
relations have been used in several different contexts, including

\begin{itemize}
\item  standard finite quantum groups (FRT formalism \cite{FRT}),
{\em i.e. }the $RTT=TTR$ relation
\[
R_{12}T_{1}T_{2}=T_{2}T_{1}R_{12}
\]
in which $R$ solves the spectral parameter-less Yang-Baxter
equation
\[
R_{12}R_{13}R_{23}=R_{23}R_{13}R_{12}
\]
and $T$ also does not depend on any spectral parameters;

\item  quantum affine algebras (RS formalism \cite{RS}), {\em i.e.\ }the $RLL$
relations (here and below $u_{\pm }\equiv u\pm \frac{\hslash c}{4}$ unless
otherwise specified)
\begin{eqnarray*}
R_{12}(u-v)L_{1}^{\pm }(u)L_{2}^{\pm }(v) &=&L_{2}^{\pm }(v)L_{1}^{\pm
}(u)R_{12}(u-v), \\
R_{12}(u_{-}-v_{+})L_{1}^{+}(u)L_{2}^{-}(v)
&=&L_{2}^{-}(v)L_{1}^{+}(u)R_{12}(u_{+}-v_{-}),
\end{eqnarray*}
in which the first equation is defined for any spectral parameter
$u$ and $v$, while the second equation is defined only for
$|u|<|v|$, and $R$ solves the spectral Yang-Baxter equation
\[
R_{12}(u-v)R_{13}(v-w)R_{23}(u-w)=R_{23}(u-w)R_{13}(v-w)R_{12}(u-v);
\]

\item  the $RLL=LLR^{\ast }$ relations \cite{Foda1,Foda2}
\begin{eqnarray*}
R_{12}(u-v)L_{1}^{\pm }(u)L_{2}^{\pm }(v) &=&L_{2}^{\pm }(v)L_{1}^{\pm
}(u)R_{12}^{\ast }(u-v), \\
R_{12}(u_{-}-v_{+})L_{1}^{+}(u)L_{2}^{-}(v)
&=&L_{2}^{-}(v)L_{1}^{+}(u)R_{12}^{\ast }(u_{+}-v_{-}),
\end{eqnarray*}
where the first equation is defined for any spectral parameter $u$
and $v$, while the second equation is defined only for $|u|<|v|$,
and $R$ and $R^{\ast }$ both solve the spectral Yang-Baxter
equation
\[
R_{12}(u-v)R_{13}(v-w)R_{23}(u-w)=R_{23}(u-w)R_{13}(v-w)R_{12}(u-v),
\]
and are related by some analytical relations (such as modular
transformations when the entries of $R$ and $R^{\ast }$ are elliptic
functions);

\item  dynamical $RLL$ relations \cite{felder1,felder2,felder3}\
\begin{eqnarray*}
R_{12}(u-v,\lambda )L_{1}^{\pm }(u,\lambda )L_{2}^{\pm }(v,\lambda )
&=&L_{2}^{\pm }(v,\lambda )L_{1}^{\pm }(u,\lambda )R_{12}(u-v,\lambda ), \\
R_{12}(u_{-}-v_{+},\lambda )L_{1}^{+}(u,\lambda )L_{2}^{-}(v,\lambda )
&=&L_{2}^{-}(v,\lambda )L_{1}^{+}(u,\lambda )R_{12}(u_{+}-v_{-},\lambda ),
\end{eqnarray*}
in which the first equation is defined for any spectral parameter $u$ and $v$%
, while the second equation is defined only for $|u|<|v|$, and $R$ solves
the dynamical spectral Yang-Baxter equation
\begin{eqnarray*}
&&R_{12}(u-v,\lambda -h^{(3)})R_{13}(v-w,\lambda )R_{23}(u-w,\lambda
-h^{(1)}) \\
&&\qquad =R_{23}(u-w,\lambda )R_{13}(v-w,\lambda -h^{(2)})R_{12}(u-v,\lambda
);
\end{eqnarray*}

\item  the combination of $RLL=LLR^{\ast }$ and the dynamical $RLL$ relations
\cite{HY,jimbot},
\begin{eqnarray*}
R_{12}(u-v,\lambda )L_{1}^{\pm }(u,\lambda )L_{2}^{\pm }(v,\lambda )
&=&L_{2}^{\pm }(v,\lambda )L_{1}^{\pm }(u,\lambda )R_{12}^{\ast
}(u-v,\lambda ), \\
R_{12}(u_{-}-v_{+},\lambda )L_{1}^{+}(u,\lambda )L_{2}^{-}(v,\lambda )
&=&L_{2}^{-}(v,\lambda )L_{1}^{+}(u,\lambda )R_{12}^{\ast
}(u_{+}-v_{-},\lambda ),
\end{eqnarray*}
where again the third relation is defined only for $|u|<|v|$.
\end{itemize}

\vspace{1pt}

Of course, this list does not exhaust all the uses of Yang-Baxter
relations in defining quantum affine algebras --- first of all
there are versions of the above formalisms corresponding to
deformations of Lie super-algebras, in which all $L$-matrices
appearing on the left-hand side have to be multiplied by a
numerical signature matrix from the right, while all $L$-matrices
appearing on the right-hand side have to be multiplied by the same
numerical signature matrix from the left \cite{YZ} --- and there
are some other rare cases in which some other strange deformed
versions of $RLL$ relations are used.

\vspace{1pt}

Despite the numerous different forms of the Yang-Baxter relations
in different quantum algebraic contexts, there are essential
factors that are common to most of these: there are alternative,
essentially Yang-Baxter relation-free definitions (or
realizations) of the same quantum algebras (if the $R$-matrices
are of the form $R_{ij}^{kl}=a_{ij}\delta _{i}^{k}\delta
_{j}^{l}+b_{ij}\delta _{i}^{l}\delta _{j}^{k}$, anyway), and every
known such algebra ({\em i.e. }one which admits any one of the
above formalisms) falls into the class of quasi-triangular
quasi-Hopf algebras -- that is, it is a certain twist of a
standard triangular quasi-Hopf algebra \cite{D4,jimbot}.

\vspace{1pt}

In practice, there are natural reasons to use many different
realizations for the same quantum algebra. For instance, to apply
these algebras in integrable models in 2D physics, the best
formalism to use is often the $RLL$-like form; to study infinite
dimensional representations, the best formalism is the so-called
Drinfeld current realization; and to study structure and
classification problems the best formalism is the Chevalley
generating relations. For these reasons the issue of the relations
between different formalisms is an important subject, and would be
by no means an easy task were it not for the famous Ding-Frenkel
theorem which establishes an explicit relationship between the
Yang-Baxter realization and the Drinfeld current realization.

\vspace{1pt}

Now our problem arises: on the one hand, in the study of
two-parameter deformations of affine Lie (super) algebras, we have
accumulated a number of different two-parameter quantum affine
(super-)algebras which are members of the so-called infinite Hopf
family of (super-)algebras \cite{IHF1,IHF2,IHF5,IHF3,IHF4} --- a
co-structure which generalizes the standard Drinfeld coproduct
\cite{D2} for standard quantum affine algebras --- and we know
nothing about the corresponding $RLL$ formalism; on the other
hand, many authors have studied two-parameter quantum affine
algebras \cite{Rossi1,Rossi2,Foda1,Foda2,HY,jimbot,KLP,Konno,YZ2},
but from the point of view of Yang-Baxter realizations only, and
the co-structures studied are standard Drinfeld twists of Hopf
algebras. Can these results be related by use of an analogue of
Ding-Frenkel theorem? That is, does there exist an $RLL$ formalism
for two-parameter quantum affine algebras with the structure of an
infinite Hopf family of (super-)algebras? How far can we go in
generalizing both the $RLL$ formalism and the Ding-Frenkel
theorem? In this
paper, we will partly address these problems by introducing a generalized $%
RLL$ relation which contains all the above listed formalisms as
special cases. We shall consider some of the necessary conditions
for such a generalized formalism to be consistent, which result in
severe restrictions on the possible choices of the $R$-matrices.
Then, we shall establish the infinite Hopf family structure over
the generalized $RLL$ formalism and study the analogue of the
Ding-Frenkel theorem in the special case of $4\times 4$ $R$
matrices. We also outline a way to construct infinitely many
commuting operators out of our generalized $RLL$ algebras and
finally discuss briefly some possible examples of our
construction.

\vspace{1pt}

\section{The generalized formalism, associativity and co-structure}

\vspace{1pt}

\subsection{The generalized $RLL$ relation}

The generalized $RLL$ formalism we shall study is

\begin{eqnarray}
R_{12}^{(i)}(u-v,\lambda )L_{1}^{\pm (i,j)}(u,\lambda )L_{2}^{\pm
(i,j)}(v,\lambda ) &=&L_{2}^{\pm (i,j)}(v,\lambda )L_{1}^{\pm
(i,j)}(u,\lambda )R_{12}^{(j)}(u-v,\lambda ),  \label{RLL1} \\
R_{12}^{(i)}(u_{-}-v_{+},\lambda )L_{1}^{+(i,j)}(u,\lambda
)L_{2}^{-(i,j)}(v,\lambda ) &=&L_{2}^{-(i,j)}(v,\lambda
)L_{1}^{+(i,j)}(u,\lambda )R_{12}^{(j)}(u_{+}-v_{-},\lambda ),\qquad |u|<|v|,
\label{RLL2} \\
u_{\pm } &\equiv &u\pm \frac{\hslash c^{(i,j)}}{4},\qquad (c^{(i,j)}\quad
\mathrm{is\quad central})  \nonumber
\end{eqnarray}

Unlike the usual $RLL$ formalisms listed in the introduction, the
$R$ matrices are now labeled by an upper index $(i)$ and the
$L$-matrices by two upper indices $(i,j)$ where $i,j\in
\mathbb{Z}$. The parameters $u,v$ are the usual spectral
parameters, and $\lambda $ is a dynamical variable (which may or
may not appear at all).

At this moment we assume no restrictions on the possible form of
the $R$-matrices, and in particular we have assigned no meaning to
the extra upper indices of $R$ and $L$ and no relationship between
the two $R$-matrices carrying different upper indices at all. Such
restrictions will emerge as we consider the associativity and
co-structure of the algebras defined by the above relation.

\vspace{1pt}

\subsection{Associativity}

\vspace{1pt}Any work on quantum algebras should ensure that the
algebra being studied is associative, and, in principle, could be
equipped with a co-structure, so that tensor products of
representations may be defined. So we shall first examine our
generalized $RLL$ relations to ensure that they really define
associative algebras. To this end, let us consider the case of
product of $3$ $L$-matrices $L_{1}^{(i,j)}(u,\lambda
)L_{2}^{(i,j)}(v,\lambda )L_{3}^{(i,j)}(w,\lambda )$. If the
algebra is associative, we should be able to apply this product in
two different ways, {\em i.e. }
\[
L_{1}^{(i,j)}(u,\lambda )L_{2}^{(i,j)}(v,\lambda )L_{3}^{(i,j)}(w,\lambda
)=\lbrack L_{1}^{(i,j)}(u,\lambda )L_{2}^{(i,j)}(v,\lambda )\rbrack
L_{3}^{(i,j)}(w,\lambda )=L_{1}^{(i,j)}(u,\lambda )\lbrack
L_{2}^{(i,j)}(v,\lambda )L_{3}^{(i,j)}(w,\lambda )\rbrack .
\]

Now temporarily we omit the upper indices $\pm $ of $L^{(i,j)}$
and arguments of $R^{(i)}$ and $L^{(i,j)}$ (though the effects of
these arguments are not neglected). We now apply the generalized
$RLL$ relations first to reverse the order of
$L_{1}^{(i,j)}L_{2}^{(i,j)}L_{3}^{(i,j)}$ and then to return to
the original order. We have
\begin{equation}
R_{32}^{(i)}R_{31}^{(i)}R_{21}^{(i)}R_{23}^{(i)}R_{13}^{(i)}R_{12}^{(i)}L_{1}^{(i,j)}L_{2}^{(i,j)}L_{3}^{(i,j)}=L_{1}^{(i,j)}L_{2}^{(i,j)}L_{3}^{(i,j)}R_{32}^{(j)}R_{31}^{(j)}R_{21}^{(j)}R_{23}^{(j)}R_{13}^{(j)}R_{12}^{(j)},
\label{asso}
\end{equation}
where the order of lower indices of the $L$'s is changed in the
following manne :
\begin{eqnarray*}
(123) &\rightarrow &(213)\rightarrow (231)\rightarrow (321) \\
&\rightarrow &(312)\rightarrow (132)\rightarrow (123).
\end{eqnarray*}
On the first line we changed $L_{1}^{(i,j)}L_{2}^{(i,j)}L_{3}^{(i,j)}$ into $%
L_{3}^{(i,j)}L_{2}^{(i,j)}L_{1}^{(i,j)}$ using left-grouping of the
successive products, while on the second line we changed $%
L_{3}^{(i,j)}L_{2}^{(i,j)}L_{1}^{(i,j)}$ back into $%
L_{1}^{(i,j)}L_{2}^{(i,j)}L_{3}^{(i,j)}$ using right-grouping, so
associativity has been implied in the above operation. A
sufficient but probably not too restrictive condition to ensure
the correctness of the equation (\ref{asso}) is that
\[
R_{32}^{(i)}R_{31}^{(i)}R_{21}^{(i)}R_{23}^{(i)}R_{13}^{(i)}R_{12}^{(i)}=R_{32}^{(j)}R_{31}^{(j)}R_{21}^{(j)}R_{23}^{(j)}R_{13}^{(j)}R_{12}^{(j)}=1.
\]
Therefore, assuming unitarity of $R$-matrices, {\em i.e. }
\begin{eqnarray}
R_{12}^{(i)}(u,\lambda )R_{21}^{(i)}(-u,\lambda ) &=&1,  \label{unitarity1}
\\
R_{12}^{(j)}(u,\lambda )R_{21}^{(j)}(-u,\lambda ) &=&1,  \label{unitarity2}
\end{eqnarray}
we conclude that $R^{(i)}$ and $R^{(j)}$ must separately satisfy the
(probably generalized) dynamical Yang-Baxter equation
\begin{eqnarray}
R_{12}^{(k)}(u-v,\lambda -h^{(3)})R_{13}^{(k)}(v-w,\lambda
)R_{23}^{(k)}(u-w,\lambda -h^{(1)})\hspace{2.6in}
&&\nonumber\\\hspace{1.6in}=\qquad R_{23}^{(k)}(u-w,\lambda
)R_{13}^{(k)}(v-w,\lambda -h^{(2)})R_{12}^{(k)}(u-v,\lambda
),\qquad(k=i,j).& \label{YBE}
\end{eqnarray}
Later we shall see that unitarity of the $R$-matrices is also
required if the Drinfeld current realization is considered, and
will introduce a coproduct which is compatible with the infinite
Hopf family in that realization.

\vspace{1pt}Let us stress that so far we have not assumed any
relationship between $R^{(i)}$ and $R^{(j)}$. The different upper
indices only indicate that the two $R$-matrices may be different.
Thus associativity has not resulted in any restrictions on the
relation between $R$-matrices.

\vspace{1pt}

\subsection{Co-structure}

\vspace{1pt}In contrast, however, the definition of a co-structure
does require some relationship between $R^{(i)}$ and $R^{(j)}$, as
we discuss below.

\vspace{1pt}

First we recall what we mean by a co-structure. Algebraically a
co-structure is a property of algebras which allows the definition
of a tensor product between two algebras (usually two copies of
the same algebra --- in that case
the algebra is called co-closed). For instance, for algebras defined by
the non-dynamical $RLL=LLR$ relations, the co-structure is
just the standard Hopf algebra structure; in particular, the
coproduct is simply an operation which creates a generalized $L$
matrix $\mathcal{L}^{(n)}$ which
obey the $R\mathcal{L}^{(n)}\mathcal{L}^{(n)}=\mathcal{L}^{(n)}\mathcal{L}%
^{(n)}R$ relation with the same $R$, where
\begin{eqnarray*}
\mathcal{L}^{(n)} &=&\Delta ^{(n)}L \\
&\equiv &(id\otimes ...\otimes id\otimes \Delta )\circ (id\otimes ...\otimes
\Delta )\circ ...\circ (id\otimes \Delta )\circ \Delta L \\
&=&L\dot{\otimes}L\dot{\otimes}...\dot{\otimes}L
\end{eqnarray*}
is the $n$-th coproduct of $L$, in which $L$ can be either $L^{+}$ or $L^{-}$,
and $L \dot{\otimes} L$ is defined via $(L \dot{\otimes} L)_b^a
=\sum_c L_c^a \otimes L_b^c$. The
point is that applying $R_{12}$ to the left hand side of $\mathcal{L}_{1}%
\mathcal{L}_{2}$ would result in
\begin{equation}
R_{12}\mathcal{L}_{1}^{(n)}\mathcal{L}_{2}^{(n)}=L_{2}L_{1}R_{12}\mathcal{L}%
_{1}^{(n-1)}\mathcal{L}_{2}^{(n-1)}=...=\mathcal{L}_{2}^{(n)}\mathcal{L}%
_{1}^{(n)}R_{12},  \label{Ln}
\end{equation}
so that the $n$-th coproduct $\Delta ^{(n)}$ is an algebra homomorphism.

For algebras defined by relations like $RLL=LLR^{\ast }$, this is impossible
because after the first use of the $RLL$ relation, a different $R$-matrix $%
R^{\ast }$ appears and the iteration stops. One possible way round
is to introduce a twist operation such that the $R^{\ast }$ which
appears after each use of the $RLL$ relation is twisted back to
$R$ and the iteration can be continued. This amounts to the notion
of the Drinfeld twist in quasi-Hopf algebras, and the possibility
of defining the twist operation imposes severe restrictions on the
relations between $R$ and $R^{\ast }$.

Since we assumed no relationship between $R^{(i)}$ and $R^{(j)}$
for $i\neq j $ so far, our situation is very much like the case of
$RLL=LLR^{\ast }$ relations. However, our aim here is not to
resort to a Drinfeld twist but rather to incorporate the structure
of the infinite Hopf family in the generalized $RLL$ formalism.
Therefore, instead of retaining the tensor product of many copies
of the same algebra, we can think about making tensor products
between different algebras defined by our generalized \ $RLL$
formalism --- this is exactly what we did in the Drinfeld current
realization of infinite Hopf family of algebras.

In order to incorporate the infinite Hopf family structure
directly at the level of $RLL$ formalism, however, we have to
introduce a relationship between the two $R$-matrices $R^{(i)}$
and $R^{(j)}$ in our generalized $RLL$ formalism.

\vspace{1pt}First we choose some unitary solution of the
(dynamical) Yang-Baxter equation (\ref{YBE}), denoted
$R^{(0)}(u,\lambda )$. We then introduce invertible operators
$\rho _{0}^{\pm }$ which act on this $R$-matrix to give $R^{(\pm
1)}(u,\lambda )=$ $\rho _{0}^{\pm
}R^{(0)}(u,\lambda )$. {\em The permitted operators $\rho _{0}^{\pm }$%
 are such that they preserve unitarity as well as the Yang-Baxter
equation}. On $R^{(\pm 1)}(u,\lambda )$ we may again act with
$\rho _{\pm 1}^{\pm }$ (with the same property as $\rho _{0}^{\pm
}$) to get $R^{(\pm 2)}(u,\lambda )$ {\em etc}. Assume that we have a
large (possibly infinite) set $%
\Upsilon $ of $R$-matrices whose elements are obtained by
recursively applying the relation $R^{\lbrack \pm (n+1)\rbrack
}(u,\lambda )\equiv (\rho
_{n}^{\pm })R^{(\pm n)}(u,\lambda )$. An additional requirement is that $%
\rho _{n+1}^{-}\circ \rho _{n}^{+}=\rho _{n-1}^{+}\circ \rho _{n}^{-}=id_{n}$%
, so that $\Upsilon $ forms a single $\rho _{n}^{\pm }$ orbit. We
may then put these $R$-matrices into our generalized $RLL$
relations.

These relations define a huge family of associative algebras $\{\mathcal{A}%
_{i,j}\equiv \mathcal{A}(R^{(i)},R^{(j)}),i,j\in \mathbb{Z}\}$ , each
labeled by two ordered integers corresponding to the upper indices of the two $%
R$-matrices in the above relations. Among these algebras we are
particularly interested in the ones labeled by two successive
integers, {\em i.e.\ }$\mathcal{A}_{n,n+1}\equiv \mathcal{A}_{n}$.
For these algebras we simplify
the notation as follows: $L^{\pm (n,n+1)}\rightarrow L^{\pm (n)}$, $%
c^{(n,n+1)}\rightarrow c_{n}$. It is remarkable that for each pair
of algebras $\mathcal{A}_{n}$ and $\mathcal{A}_{n\pm 1}$ there
exists an algebra morphism
\[
\tau _{n}^{\pm }:\mathcal{A}_{n}\rightarrow \mathcal{A}_{n\pm 1}
\]
defined via
\begin{eqnarray*}
&\tau _{n}^{+}L^{\pm (n)}=L^{\pm (n+1)},\qquad \tau _{n}^{-}L^{\pm
(n)}=L^{\pm (n-1)},& \\
&\tau _{n}^{\pm }R^{(n)}(u,\lambda )=\rho _{n}^{\pm }R^{(n)}(u,\lambda
)=R^{(n\pm 1)}(u,\lambda ),& \\
&\tau _{n}^{\pm }R^{(n\pm 1)}(u,\lambda )=\rho _{n\pm 1}^{\pm }R^{(n\pm
1)}(u,\lambda )=R^{(n\pm 2)}(u,\lambda ).&
\end{eqnarray*}
The algebra morphisms $\tau _{n}^{\pm }$ also obey the condition
\[
\tau _{n+1}^{-}\circ \tau _{n}^{+}=\tau _{n-1}^{+}\circ \tau
_{n}^{-}=id_{n},
\]
because they are essentially the lift of the action of $\rho _{n}^{\pm }$
onto the algebras $\mathcal{A}_{n}$. For any pair of integers $n<m,$ the
composition of $\tau _{n}^{\pm }$%
\begin{eqnarray}
&&Mor(\mathcal{A}_{m},~\mathcal{A}_{n})\ni \tau ^{(m,n)}\equiv \tau
_{m-1}^{+}...\tau _{n+1}^{+}\tau _{n}^{+}:~~\mathcal{A}_{n}\rightarrow
\mathcal{A}_{m},  \nonumber \\
&&Mor(\mathcal{A}_{n},~\mathcal{A}_{m})\ni \tau ^{(n,m)}\equiv \tau
_{n+1}^{-}...\tau _{m-1}^{-}\tau _{m}^{-}:~~\mathcal{A}_{m}\rightarrow
\mathcal{A}_{n}  \label{taumn}
\end{eqnarray}

\noindent are algebra morphisms between $\mathcal{A}_{n}$ and $\mathcal{A}%
_{m}$, and since $\tau ^{(m,n)}\tau ^{(n,m)}=id_{m},~\tau ^{(n,m)}\tau
^{(m,n)}=id_{n}$ and $\tau ^{(m,p)}\tau ^{(p,n)}=\tau ^{(m,n)}$, they make
the family of algebras $\{\mathcal{A}_{n},~n\in \mathbb{Z}\}$ into a
category.

Let us now define the following co-structure:

\begin{itemize}
\item  Coproduct
\begin{eqnarray*}
\Delta _{n}^{+}L^{\pm (n)}(u,\lambda ) &=&L^{\pm (n)}(u\pm \frac{\hslash
c_{n+1}}{4})\dot{\otimes}L^{\pm (n+1)}(u\mp \frac{\hslash c_{n}}{4}), \\
\Delta _{i}^{-}L^{\pm (n)}(u,\lambda ) &=&L^{\pm (n-1)}(u\pm \frac{\hslash
c_{n}}{4})\dot{\otimes}L^{\pm (n)}(u\mp \frac{\hslash c_{n-1}}{4});
\end{eqnarray*}

\item  Antipode
\begin{eqnarray*}
S_{n}^{+}L^{\pm (n)}(u,\lambda ) &=&L^{\pm (n+1)}(u,\lambda )^{-1}, \\
S_{n}^{-}L^{\pm (n)}(u,\lambda ) &=&L^{\pm (n-1)}(u,\lambda )^{-1};
\end{eqnarray*}

\item  Counit
\[
\epsilon _{n}L^{\pm (n)}(u,\lambda )=id_{n}.
\]
\end{itemize}

By straightforward verification, we find that the above
co-structure obeys the following axioms for infinite Hopf family
of algebras \cite{IHF1,IHF2,IHF3,IHF4,IHF5} :

\begin{itemize}
\item  $(\epsilon _{n}\otimes id_{n+1})\circ \Delta _{n}^{+}=\tau
_{n}^{+},~(id_{n-1}\otimes \epsilon _{n})\circ \Delta _{n}^{-}=\tau _{n}^{-}$

\item  $m_{n+1}\circ (S_{n}^{+}\otimes id_{n+1})\circ \Delta
_{n}^{+}=\epsilon _{n+1}\circ \tau _{n}^{+},~m_{n-1}\circ (id_{n-1}\otimes
S_{n}^{-})\circ \Delta _{n}^{-}=\epsilon _{n-1}\circ \tau _{n}^{-}$

\item  $(\Delta _{n}^{-}\otimes id_{n+1})\circ \Delta
_{n}^{+}=(id_{n-1}\otimes \Delta _{n}^{+})\circ \Delta _{n}^{-}$
\end{itemize}

\noindent in which $m_{n}$ is the algebra multiplication for $\mathcal{A}%
_{n} $. Moreover, it is easy to check that the images of $\Delta
_{n}^{\pm }$
satisfy the defining relations of $\mathcal{A}_{n,n+2}$ with the center $%
c_{n}+c_{n+1}$, and hence $\Delta _{n}^{\pm }$ is an algebra
homomorphism. By using compositions of the coproduct, $\Delta
_{n}^{(m)+}\equiv (id_{n}\otimes id_{n+1}\otimes ...\otimes
id_{n+m-2}\otimes \Delta _{n+m-1}^{+})\circ (id_{n}\otimes
id_{n+1}\otimes ...\otimes id_{n+m-3}\otimes \Delta
_{n+m-2}^{+})...\circ (id_{n}\otimes \Delta
_{n+1}^{+})\circ \Delta _{n}^{+}$ , we can show that all the algebras $%
\mathcal{A}_{n,m}$ are inter-related by the co-structure. Indeed, denoting
\[
\mathcal{L}^{\pm (n,m)}(u,\lambda )\equiv \Delta _{n}^{(m)+}L^{\pm
(n)}(u,\lambda ),
\]
one can check that the following relations hold:
\begin{eqnarray}
R_{12}^{(n)}(u-v,\lambda )\mathcal{L}_{1}^{\pm (n,m)}(u,\lambda )\mathcal{L}%
_{2}^{\pm (n,m)}(v,\lambda ) &=&\mathcal{L}_{2}^{\pm (n,m)}(v,\lambda )%
\mathcal{L}_{1}^{\pm (n,m)}(u,\lambda )R_{12}^{(m)}(u-v,\lambda ),
\label{Lpmn1} \\
R_{12}^{(n)}(u_{-}-v_{+},\lambda )\mathcal{L}_{1}^{+(n,m)}(u,\lambda )%
\mathcal{L}_{2}^{-(n,m)}(v,\lambda ) &=&\mathcal{L}_{2}^{-(n,m)}(v,\lambda )%
\mathcal{L}_{1}^{+(n,m)}(u,\lambda )R_{12}^{(m)}(u_{+}-v_{-},\lambda
),\qquad |u|<|v|,  \label{Lpmn2} \\
u_{\pm } &\equiv &u\pm \frac{\hslash (c_{n}+c_{n+1}+...+c_{m-1})}{4}.
\nonumber
\end{eqnarray}
These relations are exactly the generating relations of
$\mathcal{A}_{n,m}$ with the center $c^{(n,m)}$ given by
$c^{(n,m)}=c_{n}+c_{n+1}+...+c_{m-1}$. Equations (\ref{Lpmn1}) and
(\ref{Lpmn2}) are the analogues of (\ref{Ln}) in the case of
infinite Hopf family of algebras.

\vspace{1pt}

\textit{Remark}. The co-structure of an infinite Hopf family of
algebras may look quite unnatural at first sight to readers who
are familiar with standard Hopf algebra co-structures, because it
makes use of tensor products between different algebras. However,
the idea of making tensor product between different algebras is
actually not so surprising --- it has already been studied in
several different contexts \cite{cross1,cross2,cross3,cross4}
which are not related to our purpose.

\section{Gauss Decomposition}

\vspace{1pt} In this section, we assume that the $R$-matrices have
the form
\begin{equation}
\lbrack R^{(n)}\rbrack _{ij}^{kl}(u-v,\lambda )=a_{ij}^{(n)}(u-v,\lambda
)\delta _{i}^{k}\delta _{j}^{l}+b_{ij}^{(n)}(u-v,\eta ,\lambda )\delta
_{i}^{l}\delta _{j}^{k}.  \label{Rform}
\end{equation}
This is because it is known that for $R$-matrices of more complex
form, there is no analogue of the Ding-Frenkel theorem (such
``bad'' cases include the $8$-vertex $R$-matrix
\cite{Foda1,Foda2}).

The proof of the Ding-Frenkel theorem for $R$-matrices of
arbitrary size is a tedious and complicated piece of work, and
should in general be accomplished by induction over the size of
$R$-matrices \cite{DF}. In this paper, however, we shall not make
the induction over the size of $R$-matrices, but rather will
illustrate our generalized $RLL$ formalism only in the simplest
case of $4\times 4$ $R$-matrices.

\vspace{1pt}Before we proceed, let us remark that the construction
of the last section works fine if we understand all the products
between $R$ and $L$ and those among $L$s as graded matrix products
for super matrices, and simultaneously understand all the tensor
products as graded tensor products. In this way we would have the
$RLL$ formalism for an infinite Hopf family of super-algebras. In
practice, however, it is usually better to change the graded
matrix product into the usual matrix products. For members of the
infinite Hopf family of super-algebras, this amounts to altering
the $RLL$ relations into the form
\begin{eqnarray}
R_{12}^{(i)}(u-v,\lambda )L_{1}^{\pm (i,j)}(u,\lambda )\varpi
L_{2}^{\pm (i,j)}(v,\lambda )\varpi &=&\varpi L_{2}^{\pm
(i,j)}(v,\lambda )\varpi L_{1}^{\pm (i,j)}(u,\lambda
)R_{12}^{(j)}(u-v,\lambda ),  \label{RLLO1} \\
R_{12}^{(i)}(u_{-}-v_{+},\lambda )L_{1}^{+(i,j)}(u,\lambda )\varpi
L_{2}^{-(i,j)}(v,\lambda )\varpi &=&\varpi
L_{2}^{-(i,j)}(v,\lambda )\varpi L_{1}^{+(i,j)}(u,\lambda
)R_{12}^{(j)}(u_{+}-v_{-},\lambda ),\hspace{4pt} |u|<|v|
\;\;\label{RLLO2} \\ u_{\pm } &\equiv &u\pm \frac{\hslash
c^{(i,j)}}{4},\qquad (c^{(i,j)}\quad \mathrm{is\quad central})
\nonumber
\end{eqnarray}
where $\varpi $ is a diagonal numerical matrix given by $\varpi
_{ij}^{kl}=(-1)^{\lbrack i\rbrack \lbrack j\rbrack }\delta
_{i}^{k}\delta _{j}^{l}$ which reflects the grading of the $R$ and
$L$-matrices. We may also write the usual (non-super) $RLL$
relations in the above form, provided we understand $\varpi $ as
the identity matrix. This section will be based on $RLL$ relations
of the above form, putting the deformations corresponding to the
usual and super root systems on an equal footing.

The most general $4\times 4$ $R$-matrix of the form (\ref{Rform})
is written
\begin{equation}
R^{(i)}(u,\lambda )=\left(
\begin{array}{llll}
a^{(i)}(u,\lambda ) & 0 & 0 & 0 \\
0 & b^{(i)}(u,\lambda ) & t^{(i)}(u,\lambda ) & 0 \\
0 & s^{(i)}(u,\lambda ) & c^{(i)}(u,\lambda ) & 0 \\
0 & 0 & 0 & d^{(i)}(u,\lambda )
\end{array}
\right)  \label{R4}
\end{equation}
The corresponding numerical signature matrix $\varpi $ is given by
\[
\varpi =\left(
\begin{array}{llll}
1 &  &  &  \\
& 1 &  &  \\
&  & 1 &  \\
&  &  & \epsilon
\end{array}
\right) ,\qquad \epsilon =\pm 1.
\]
In such cases the $L$-matrices can be represented as operator-valued $%
2\times 2$ matrices and we adopt the following Gauss decompositions (we omit
the upper indices $(i,j)$ here and below):
\begin{eqnarray*}
L^{\pm }(u,\lambda ) &=&\left(
\begin{array}{ll}
1 & 0 \\
f^{\pm }(u,\lambda ) & 1
\end{array}
\right) \left(
\begin{array}{ll}
k_{1}^{\pm }(u,\lambda ) & 0 \\
0 & k_{2}^{\pm }(u,\lambda )
\end{array}
\right) \left(
\begin{array}{ll}
1 & e^{\pm }(u,\lambda ) \\
0 & 1
\end{array}
\right) \\
&=&\left(
\begin{array}{ll}
k_{1}^{\pm }(u,\lambda ) & k_{1}^{\pm }(u,\lambda )e^{\pm }(u,\lambda ) \\
f^{\pm }(u,\lambda )k_{1}^{\pm }(u,\lambda ) & k_{2}^{\pm }(u,\lambda
)+f^{\pm }(u,\lambda )k_{1}^{\pm }(u,\lambda )e^{\pm }(u,\lambda )
\end{array}
\right) ,
\end{eqnarray*}
\begin{eqnarray*}
L^{\pm }(u,\lambda )^{-1} &=&\left(
\begin{array}{ll}
1 & -e^{\pm }(u,\lambda ) \\
0 & 1
\end{array}
\right) \left(
\begin{array}{ll}
k_{1}^{\pm }(u,\lambda )^{-1} & 0 \\
0 & k_{2}^{\pm }(u,\lambda )^{-1}
\end{array}
\right) \left(
\begin{array}{ll}
1 & 0 \\
-f^{\pm }(u,\lambda ) & 1
\end{array}
\right) \\
&=&\left(
\begin{array}{ll}
k_{1}^{\pm }(u,\lambda )^{-1}+e^{\pm }(u,\lambda )k_{2}^{\pm }(u,\lambda
)^{-1}f^{\pm }(u,\lambda ) & -e^{\pm }(u,\lambda )k_{2}^{\pm }(u,\lambda
)^{-1} \\
-k_{2}^{\pm }(u,\lambda )^{-1}f^{\pm }(u,\lambda ) & k_{2}^{\pm }(u,\lambda
)^{-1}
\end{array}
\right) .
\end{eqnarray*}
For simplicity of notation, we also denote $k_{i}^{\pm }(u),$
$e^{\pm }(u)$ and $f^{\pm }(u)$ without explicitly referring to
the dependence on $\lambda $. This dependence, however, is always
implicitly assumed.

\vspace{1pt}

To get the desired analogue of the Ding-Frenkel theorem, it is
convenient to rewrite the $RLL$ relations in several equivalent
ways. These include
\begin{eqnarray}
L_{2}^{\pm }(v,\lambda )^{-1}\varpi R_{12}^{(i)}(u-v,\lambda )L_{1}^{\pm
}(u,\lambda )\varpi &=&\varpi L_{1}^{\pm }(u,\lambda
)R_{12}^{(j)}(u-v,\lambda )\varpi L_{2}^{\pm }(v,\lambda )^{-1},  \nonumber
\\
L_{2}^{-}(v,\lambda )^{-1}\varpi R_{12}^{(i)}(u_{-}-w_{+},\lambda
)L_{1}^{+}(u,\lambda )\varpi &=&\varpi L_{1}^{+}(u,\lambda
)R_{12}^{(j)}(u_{+}-v_{-},\lambda )\varpi L_{2}^{-}(v,\lambda )^{-1};
\label{II}
\end{eqnarray}
\begin{eqnarray}
L_{1}^{\pm }(u,\lambda )^{-1}\varpi L_{2}^{\pm }(v,\lambda )^{-1}\varpi
R_{12}^{(i)}(u-v,\lambda ) &=&R_{12}^{(j)}(u-v,\lambda )\varpi L_{2}^{\pm
}(v,\lambda )^{-1}\varpi L_{1}^{\pm }(u,\lambda )^{-1},  \nonumber \\
L_{1}^{+}(u,\lambda )^{-1}\varpi L_{2}^{-}(v,\lambda )^{-1}\varpi
R_{12}^{(i)}(u_{-}-w_{+},\lambda ) &=&R_{12}^{(j)}(u_{+}-v_{-},\lambda
)\varpi L_{2}^{-}(v,\lambda )^{-1}\varpi L_{1}^{+}(u,\lambda )^{-1};
\label{III}
\end{eqnarray}
\begin{eqnarray}
L_{1}^{\pm }(u,\lambda )^{-1}R_{21}^{(i)}(v-u,\lambda )\varpi L_{2}^{\pm
}(v,\lambda )\varpi &=&\varpi L_{2}^{\pm }(v,\lambda )\varpi
R_{21}^{(j)}(v-u,\lambda )L_{1}^{\pm }(u,\lambda )^{-1},  \nonumber \\
L_{1}^{+}(u,\lambda )^{-1}R_{21}^{(i)}(v_{+}-u_{-},\lambda )\varpi
L_{2}^{-}(v,\lambda )\varpi &=&\varpi L_{2}^{-}(v,\lambda )\varpi
R_{21}^{(j)}(v_{-}-u_{+},\lambda )L_{1}^{+}(u,\lambda )^{-1},  \label{IV}
\end{eqnarray}
where the unitarity of the $R$-matrices is implied by the
equivalence of these different equations.

\vspace{1pt}

Expanding the above equations into matrix components, we get, after some
algebra,\vspace{1pt} the following relations:

\begin{enumerate}
\item  Relations among $k_{i}^{\pm }(u)$:
\begin{eqnarray}
a^{(i)}(u-v,\lambda )k_{1}^{\pm }(u)k_{1}^{\pm }(v) &=&k_{1}^{\pm
}(v)k_{1}^{\pm }(u)a^{(j)}(u-v,\lambda ),  \label{kk1} \\
a^{(i)}(u_{-}-v_{+},\lambda )k_{1}^{+}(u)k_{1}^{-}(v)
&=&k_{1}^{-}(v)k_{1}^{+}(u)a^{(j)}(u_{+}-v_{-},\lambda ),  \label{kk2} \\
k_{2}^{\pm }(v)^{-1}b^{(i)}(u-v,\lambda )k_{1}^{\pm }(u) &=&k_{1}^{\pm
}(u)b^{(j)}(u-v,\lambda )k_{2}^{\pm }(v)^{-1},  \label{kk3} \\
k_{2}^{-}(v)^{-1}b^{(i)}(u_{-}-v_{+},\lambda )k_{1}^{+}(u)
&=&k_{1}^{+}(u)b^{(j)}(u_{+}-v_{-},\lambda )k_{2}^{-}(v)^{-1},  \label{kk4}
\\
k_{2}^{\pm }(u)^{-1}k_{2}^{\pm }(v)^{-1}d^{(i)}(u-v,\lambda )
&=&d^{(j)}(u-v,\lambda )k_{2}^{\pm }(v)^{-1}k_{2}^{\pm }(u)^{-1},
\label{kk5} \\
k_{2}^{+}(u)^{-1}k_{2}^{-}(v)^{-1}d^{(i)}(u_{-}-v_{+},\lambda )
&=&d^{(j)}(u_{+}-v_{-},\lambda )k_{2}^{-}(v)^{-1}k_{2}^{+}(u)^{-1},
\label{kk6} \\
k_{2}^{\pm }(u)^{-1}b^{(i)}(v-u,\lambda )k_{1}^{\pm }(v) &=&k_{1}^{\pm
}(v)b^{(j)}(v-u,\lambda )k_{2}^{\pm }(u)^{-1},  \label{kk7} \\
k_{2}^{+}(u)^{-1}b^{(i)}(v_{+}-u_{-},\lambda )k_{1}^{-}(v)
&=&k_{1}^{-}(v)b^{(j)}(v_{-}-u_{+},\lambda )k_{2}^{+}(u)^{-1},  \label{kk8}
\end{eqnarray}

\item  relations between $k_{i}^{\pm }(u)$ and $e^{\pm }(v)$, $f^{\pm }(v)$:
\begin{eqnarray*}
&k_{1}^{\pm }(u)a^{(j)}(u-v,\lambda )e^{\pm }(v)-e^{\pm }(v)k_{1}^{\pm
}(u)b^{(j)}(u-v,\lambda )-k_{1}^{\pm }(u)e^{\pm }(u)s^{(j)}(u-v,\lambda )=0,&
\\
&b^{(i)}(u-v,\lambda )k_{1}^{\pm }(u)f^{\pm }(v)+t^{(i)}(u-v,\lambda )f^{\pm
}(u)k_{1}^{\pm }(u)-f^{\pm }(v)a^{(i)}(u-v,\lambda )k_{1}^{\pm }(u)=0;& \\
&k_{1}^{+}(u)a^{(j)}(u_{+}-v_{-},\lambda
)e^{-}(v)-e^{-}(v)k_{1}^{+}(u)b^{(j)}(u_{+}-v_{-},\lambda
)-k_{1}^{+}(u)e^{+}(u)s^{(j)}(u_{+}-v_{-},\lambda )=0,& \\
&b^{(i)}(u_{-}-v_{+},\lambda
)k_{1}^{+}(u)f^{-}(v)+t^{(i)}(u_{-}-v_{+},\lambda
)f^{+}(u)k_{1}^{+}(u)-f^{-}(v)a^{(i)}(u_{-}-v_{+},\lambda )k_{1}^{+}(u)=0;&
\\
&k_{2}^{\pm }(v)^{-1}d^{(i)}(u-v,\lambda )f^{\pm }(u)-f^{\pm }(u)k_{2}^{\pm
}(v)^{-1}b^{(i)}(u-v,\lambda )-\epsilon k_{2}^{\pm }(v)^{-1}f^{\pm
}(v)s^{(i)}(u-v,\lambda )=0,& \\
&k_{2}^{-}(v)^{-1}d^{(i)}(u_{-}-v_{+},\lambda
)f^{+}(u)-f^{+}(u)k_{2}^{-}(v)^{-1}b^{(i)}(u_{-}-v_{+},\lambda )-\epsilon
k_{2}^{-}(v)^{-1}f^{-}(v)s^{(i)}(u_{-}-v_{+},\lambda )=0;& \\
&b^{(j)}(u-v,\lambda )k_{2}^{\pm }(v)^{-1}e^{\pm }(u)-e^{\pm
}(u)d^{(j)}(u-v,\lambda )k_{2}^{\pm }(v)^{-1}+\epsilon t^{(j)}(u-v,\lambda
)e^{\pm }(v)k_{2}^{\pm }(v)^{-1}=0,& \\
&b^{(j)}(u_{+}-v_{-},\lambda
)k_{2}^{-}(v)^{-1}e^{+}(u)-e^{+}(u)d^{(j)}(u_{+}-v_{-},\lambda
)k_{2}^{-}(v)^{-1}+\epsilon t^{(j)}(u_{+}-v_{-},\lambda
)e^{-}(v)k_{2}^{-}(v)^{-1}=0;& \\
&b^{(i)}(v_{+}-u_{-},\lambda
)k_{1}^{-}(v)f^{+}(u)-f^{+}(u)a^{(i)}(v_{+}-u_{-},\lambda
)k_{1}^{-}(v)+t^{(i)}(v_{+}-u_{-},\lambda )f^{-}(v)k_{1}^{-}(v)=0,& \\
&e^{+}(u)k_{1}^{-}(v)b^{(j)}(v_{-}-u_{+},\lambda
)-k_{1}^{-}(v)a^{(j)}(v_{-}-u_{+},\lambda
)e^{+}(u)+k_{1}^{-}(v)e^{-}(v)s^{(j)}(v_{-}-u_{+},\lambda )=0;& \\
&e^{-}(v)d^{(j)}(v_{-}-u_{+},\lambda )k_{2}^{+}(u)^{-1}-\epsilon
t^{(j)}(v_{-}-u_{+},\lambda
)e^{+}(u)k_{2}^{+}(u)^{-1}-b^{(j)}(v_{-}-u_{+},\lambda
)k_{2}^{+}(u)^{-1}e^{-}(v)=0,& \\
&f^{-}(v)k_{2}^{+}(u)^{-1}b^{(i)}(v_{+}-u_{-},\lambda
)-k_{2}^{+}(u)^{-1}d^{(i)}(v_{+}-u_{-},\lambda )f^{-}(v)+\epsilon
k_{2}^{+}(u)^{-1}f^{+}(u)s^{(i)}(v_{+}-u_{-},\lambda )=0;&
\end{eqnarray*}

\item  Relations containing two $e^{\pm }$'s or two $f^{\pm }$'s:
\begin{eqnarray*}
&a^{(i)}(u-v,\lambda )k_{1}^{\pm }(u)e^{\pm }(u)k_{1}^{\pm }(v)e^{\pm
}(v)-\epsilon k_{1}^{\pm }(v)e^{\pm }(v)k_{1}^{\pm }(u)e^{\pm
}(u)d^{(j)}(u-v,\lambda )=0,& \\
&a^{(i)}(u_{-}-v_{+},\lambda
)k_{1}^{+}(u)e^{+}(u)k_{1}^{-}(v)e^{-}(v)-\epsilon
k_{1}^{-}(v)e^{-}(v)k_{1}^{+}(u)e^{+}(u)d^{(j)}(u_{+}-v_{-},\lambda )=0,& \\
&d^{(i)}(u-v,\lambda )f^{\pm }(u)k_{1}^{\pm }(u)f^{\pm }(v)k_{1}^{\pm
}(v)-\epsilon f^{\pm }(v)k_{1}^{\pm }(v)f^{\pm }(u)k_{1}^{\pm
}(u)a^{(j)}(u-v,\lambda )=0,& \\
&d^{(i)}(u_{-}-v_{+},\lambda
)f^{+}(u)k_{1}^{+}(u)f^{-}(v)k_{1}^{-}(v)-\epsilon
f^{-}(v)k_{1}^{-}(v)f^{+}(u)k_{1}^{+}(u)a^{(j)}(u_{+}-v_{-},\lambda )=0;&
\end{eqnarray*}

\item  Mixed relations between $e^{\pm }$ and $f^{\pm }$'s:
\begin{eqnarray}
-\epsilon e^{\pm }(u)f^{\pm }(v)+f^{\pm }(v)e^{\pm }(u) &=&k_{1}^{\pm
}(u)^{-1}b^{(i)}(u-v,\lambda )^{-1}t^{(i)}(u-v,\lambda )k_{2}^{\pm }(u)
\nonumber \\
&&-k_{2}^{\pm }(v)b^{(j)}(u-v,\lambda )^{-1}t^{(j)}(u-v,\lambda )k_{1}^{\pm
}(v)^{-1},  \label{pm}
\end{eqnarray}
\begin{eqnarray}
-\epsilon e^{+}(u)f^{-}(v)+f^{-}(v)e^{+}(u)
&=&k_{1}^{+}(u)^{-1}b^{(i)}(u_{-}-v_{+},\lambda
)^{-1}t^{(i)}(u_{-}-v_{+},\lambda )k_{2}^{+}(u)  \nonumber \\
&&-k_{2}^{-}(v)b^{(j)}(u_{+}-v_{-},\lambda )^{-1}t^{(j)}(u_{+}-v_{-},\lambda
)k_{1}^{-}(v)^{-1},  \label{p}
\end{eqnarray}
\begin{eqnarray}
-\epsilon e^{-}(u)f^{+}(v)+f^{+}(v)e^{-}(u)
&=&k_{1}^{-}(u)^{-1}b^{(i)}(u_{+}-v_{-},\lambda
)^{-1}t^{(i)}(u_{+}-v_{-},\lambda )k_{2}^{-}(u)  \nonumber \\
&&-k_{2}^{+}(v)b^{(j)}(u_{-}-v_{+},\lambda )^{-1}t^{(j)}(u_{-}-v_{+},\lambda
)k_{1}^{+}(v)^{-1},  \label{m}
\end{eqnarray}
where the equations in (\ref{pm}) are defined for both $|u|>|v|$ and $|u|<|v|
$, equation (\ref{p}) is defined only for $|u|>|v|$ and equation (\ref{m})
is defined only for $|u|<|v|$.
\end{enumerate}

The appearance of these can be drastically simplified by defining
\begin{eqnarray*}
E(u) &=&e^{+}(u_{-})-e^{-}(u_{+}),  \label{Xp} \\
F(u) &=&f^{+}(u_{+})-f^{-}(u_{-}),  \label{Xm}
\end{eqnarray*}
giving

\begin{itemize}
\item  Relations between the $k_{i}^{\pm }$'s and the $E$ and $F$:
\begin{eqnarray}
k_{1}^{+}(u)a^{(j)}(u-v_{-},\lambda )E(v)
&=&E(v)k_{1}^{+}(u)b^{(j)}(u-v_{-},\lambda ),  \label{ke1} \\
b^{(i)}(u-v_{+},\lambda )k_{1}^{+}(u)F(v) &=&F(v)a^{(i)}(u-v_{+},\lambda
)k_{1}^{+}(u),  \label{ke2}
\end{eqnarray}
\begin{eqnarray}
E(u)k_{1}^{-}(v)b^{(j)}(v_{-}-u,\lambda )
&=&k_{1}^{-}(v)a^{(j)}(v_{-}-u,\lambda )E(u),  \label{ke3} \\
b^{(i)}(v_{+}-u,\lambda )k_{1}^{-}(v)F(u) &=&F(u)a^{(i)}(v_{+}-u,\lambda
)k_{1}^{-}(v),  \label{ke4}
\end{eqnarray}
\begin{eqnarray}
b^{(j)}(v_{-}-u,\lambda )k_{2}^{+}(u)^{-1}E(v)
&=&E(v)d^{(j)}(v_{-}-u,\lambda )k_{2}^{+}(u)^{-1},  \label{kf1} \\
k_{2}^{+}(u)^{-1}d^{(i)}(v_{+}-u,\lambda )F(v)
&=&F(v)k_{2}^{+}(u)^{-1}b^{(i)}(v_{+}-u,\lambda ).  \label{kf2}
\end{eqnarray}
\begin{eqnarray}
b^{(j)}(u-v_{-},\lambda )k_{2}^{-}(v)^{-1}E(u)
&=&E(u)d^{(j)}(u-v_{-},\lambda )k_{2}^{-}(v)^{-1},  \label{kf3} \\
k_{2}^{-}(v)^{-1}d^{(i)}(u-v_{+},\lambda )F(u)
&=&F(u)k_{2}^{-}(v)^{-1}b^{(i)}(u-v_{+},\lambda ).  \label{kf4}
\end{eqnarray}

\item  \vspace{1pt}Relations of the kind $EE$ and $FF$:
\begin{eqnarray}
a^{(j)}(v-u,\lambda )E(u)b^{(j)}(v-u,\lambda )^{-1}E(v)-\epsilon
E(v)b^{(j)}(u-v,\lambda )^{-1}E(u)d^{(j)}(u-v,\lambda ) &=&0,  \label{ee} \\
d^{(i)}(u-v,\lambda )F(u)b^{(i)}(u-v,\lambda )^{-1}F(v)-\epsilon
F(v)b^{(i)}(v-u,\lambda )^{-1}F(u)a^{(i)}(v-u,\lambda ) &=&0;  \label{ff}
\end{eqnarray}

\item  Exchange relation between $E$ and $F$ :
\begin{eqnarray}
\lbrack E(u),F(v)\rbrack _{\epsilon } &=&k_{2}^{-}(v_{-})\lbrack \Phi
^{+(j)}(u_{+}-v_{-},\lambda )-\Phi ^{-(j)}(u_{+}-v_{-},\lambda )\rbrack
k_{1}^{-}(v_{-})^{-1}  \nonumber \\
&&-k_{2}^{+}(v_{+})\lbrack \Phi ^{+(j)}(u_{-}-v_{+},\lambda )-\Phi
^{-(j)}(u_{-}-v_{+},\lambda )\rbrack k_{1}^{+}(v_{+})^{-1},  \label{EXC}
\end{eqnarray}
where $\lbrack E(u),F(v)\rbrack _{\epsilon }=E(u)F(v)-\epsilon F(v)E(u)$ and
$\Phi ^{\pm (i)}(u)$ are defined via
\[
\Phi ^{\pm (i)}(u-v,\lambda )\equiv b^{(i)}(u-v,\lambda
)^{-1}t^{(i)}(u-v,\lambda )\qquad \mathrm{for\qquad }\left\{
\begin{array}{l}
|u|>|v| \\
|u|<|v|
\end{array}
\right. .
\]
\end{itemize}

Notice that equation (\ref{EXC}) can only be understood as an
analytical continuation, because $\Phi ^{\pm (i)}(u)$ are never
simultaneously well-defined for the same $u$. In the sense of
analytical continuations, $\Phi ^{-(i)}(u)$ can be regarded as the
same as $\Phi ^{+(i)}(u)$ for almost any value of $u$ except
$u=0$, where there is a singularity. Therefore, it is convenient
to express the difference between $\Phi ^{+(i)}(u)$ and $\Phi
^{-(i)}(u)$ in terms of Dirac delta function:
\[
\Phi ^{+(i)}(u)-\Phi ^{-(i)}(u)=N^{(i)}(\hbar ,\lambda )\delta (u),
\]
where $N^{(i)}(\hbar ,\lambda )$ is a normalization coefficient which may
depend on the dynamical variable $\lambda $.

Using this notation, the exchange relation between $E(u)$ and $F(v)$ can be
rewritten as
\begin{equation}
\lbrack E(u),F(v)\rbrack _{\epsilon }=\delta
(u_{+}-v_{-})k_{2}^{-}(v_{-})N^{(j)}(\hbar ,\lambda
)k_{1}^{-}(v_{-})^{-1}-\delta (u_{-}-v_{+})k_{2}^{+}(v_{+})N^{(j)}(\hbar
,\lambda )k_{1}^{+}(v_{+})^{-1}.  \label{EXC2}
\end{equation}

Summarizing the above results, we conclude that the generating
relations for the algebra defined by our generalized $RLL$
relations (\ref{RLLO1}-{\ref {RLLO2}) with the $4\times 4$
$R$-matrices (\ref{R4}) can be written in terms of the Drinfeld
currents $k_{1,2}^{\pm }(u)$ and $E(u),$ $F(u)$ via equations
(\ref{kk1}-\ref{kk8}), (\ref{ke1}-\ref{kf4}), (\ref{ee}-\ref{ff})
and (\ref{EXC2}), provided the $R$-matrices (\ref{R4}) satisfy the
(dynamical) Yang-Baxter equation (\ref{YBE}) and the unitarity conditions (%
\ref{unitarity1}-\ref{unitarity2}). }

Let us stress that in the above calculations we made no use of
explicit formulae for the $R$-matrix entries; the only condition
used is unitarity. We also assumed no explicit commutation
relations between entries of the (dynamical) $R$-matrices and the
Drinfeld currents, so our result should hold for all unitary
dynamical $R$-matrices. Such a result is very useful when we
consider concrete examples of the generalized $%
RLL$ algebras --- to get the Drinfeld current realization, for
example, the only thing we need to do is to substitute the
concrete $R$-matrix entries into the equations
(\ref{kk1}-\ref{kk8}), (\ref{ke1}-\ref{kf4}), (\ref{ee}-\ref{ff})
and (\ref{EXC2}).

Now let us make some more comments about the essential role of the
unitarity conditions (\ref{unitarity1}-\ref{unitarity2}). In
section two, these conditions are assumed only because they are
sufficient to make the generalized $RLL$ relations associative. In
this section, however, we see that to have the Drinfeld current
realization, we need these conditions to hold. Actually there are
other reasons to impose the unitarity conditions on the
$R$-matrices. For instance, let us consider the cases when the
$R$-matrices are non-dynamical and $\epsilon =+1$. In such cases,
the equations (%
\ref{ee}) and (\ref{ff}) can be written as
\begin{eqnarray}
E(u)E(v) &=&\frac{b^{(j)}(v-u)d^{(j)}(u-v)}{a^{(j)}(v-u)b^{(j)}(u-v)}%
E(v)E(u),  \label{eeco} \\
F(u)F(v) &=&\frac{a^{(i)}(v-u)b^{(i)}(u-v)}{b^{(i)}(v-u)d^{(i)}(u-v)}%
F(v)F(u).  \label{ffco}
\end{eqnarray}

It is proven in \cite{IHF3} that the co-structure of the Drinfeld
current realization of the infinite Hopf family of algebras is
characterized solely by the structure functions appearing in the
commutation relations
\begin{eqnarray*}
E_{i}(u)E_{j}(v) &=&\Psi _{ij}(u-v|q)E_{j}(v)E_{i}(u), \\
F_{i}(u)F_{j}(v) &=&\Psi _{ij}(u-v|\tilde{q})^{-1}F_{j}(v)F_{i}(u),
\end{eqnarray*}
and the only condition that these structure functions have to obey is
\[
\Psi _{ij}(u|q)=\Psi _{ji}(-u|q)^{-1}.
\]
Now looking at the equations (\ref{eeco}) and (\ref{ffco}) we find
that the condition imposed above on $\Psi _{ij}(u|q)$ is just
\[
\frac{b^{(j)}(-u)d^{(j)}(u)}{a^{(j)}(-u)b^{(j)}(u)}=\frac{%
a^{(j)}(u)b^{(j)}(-u)}{b^{(j)}(u)d^{(j)}(-u)},
\]
which holds trivially if the unitarity condition for $%
R^{(j)}$ is satisfied. Thus the unitarity condition is not merely
among the sufficient conditions for the $RLL$ relations to be
associative, but also necessary for the Drinfeld current
realization.

\section{Comorphisms and infinitely many commuting operators}

\label{S3}

In applications of ordinary Hopf algebras, the image of the $n$-th
coproduct $\Delta^{(n)}$ is the building block for constructing
infinitely many commuting
operators in integrable/solvable models. Since $R(u-v)\mathcal{L}%
^{(n)}(u) \mathcal{L}^{(n)}(v)=\mathcal{L}^{(n)}(v) \mathcal{L}^{(n)}(u)
R(u-v)$, one can easily see that $[\mathrm{tr}\mathcal{L}^{(n)}(u), \mathrm{%
tr}\mathcal{L}^{(n)}(v)]=0$, which can be subsequently expanded
over the spectral parameter to yield infinitely many commuting
operators.

For the generalized $RLL$ algebras given by
(\ref{RLLO1}-\ref{RLLO2}), however, we cannot apply the simple
scenario above because the $R$ matrices on the two sides of
eqs.(\ref{Lpmn1}-\ref{Lpmn2}) are different. In order to get
infinitely many commuting operators associated with the
generalized $RLL$ algebras, we have to look at some new algebraic
structures which we call comorphisms, the analogue of comodules
\cite{SK2} for standard Hopf and quasi-Hopf algebras.

Throughout this section, $L^{(i,j)}$ can be either $L^{+(i,j)}$ or $%
L^{-(i,j)}$.

Now let $\{\mathcal{F}^{(n)},n\in \mathbb{Z}\}$ be another family
of algebras associated with the same $R$-matrices used in
(\ref{RLLO1}-\ref {RLLO2}), whose member $\mathcal{F}^{(i)}$ is
defined by the relation
\begin{equation}
R^{(i)}(u-v,\lambda )X_{1}^{(i)}(u,\lambda )X_{2}^{(i)}(v,\lambda
)=X_{2}^{(i)}(v,\lambda )X_{1}^{(i)}(u,\lambda ).  \label{RXX}
\end{equation}
Unlike the case of $L^{(i,j)}(u,\lambda )=(L^{(i,j)}(u,\lambda )_{a}^{b})$
which are matrices, now $X^{(i)}(u,\lambda )$ are only vectors with
components labeled by one index, $X^{(i)}(u,\lambda )=(X^{(i)}(u,\lambda
)^{a})$. Clearly the action of the operators $\rho _{n}^{\pm }$ can also be
lifted to the algebras $\mathcal{F}^{(n)}$ to give algebra morphisms. We
denote the lifted action of $\rho _{n}^{\pm }$ on $\mathcal{F}^{(n)}$by $%
\kappa _{n}^{\pm }$,
\begin{eqnarray*}
\kappa _{n}^{\pm }:\mathcal{F}^{(n)} &\rightarrow &\mathcal{F}^{(n\pm 1)}, \\
X^{(n)}(u,\lambda ) &\mapsto &X^{(n+1)}(u,\lambda ), \\
R^{(n)}(u-v,\lambda ) &\mapsto &\rho _{n}^{\pm }R^{(n)}(u-v,\lambda ).
\end{eqnarray*}

Let
\begin{eqnarray*}
\varphi ^{(n)}:\mathcal{F}^{(n)} &\rightarrow &\mathcal{A}_{n}\otimes
\mathcal{F}^{(n+1)}, \\
X^{(n)}(u,\lambda ) &\mapsto &L^{(n)}(u,\lambda )\dot{\otimes}%
X^{(n+1)}(u,\lambda )
\end{eqnarray*}
(where $(L^{(n)}(u,\lambda )\dot{\otimes}X^{(n+1)}(u,\lambda ))^{a}\equiv
\sum_{b}L^{(n)}(u,\lambda )_{b}^{a}\otimes X^{(n+1)}(u,\lambda )^{b}$) be
algebra \emph{comorphisms} in the sense that there is an algebra
homomorphism $\phi ^{(n,n+1)}:\mathcal{A}_{n}\otimes \mathcal{F}%
^{(n+1)}\rightarrow \mathcal{F}^{(n+1)}$ induced by $\varphi ^{(n)}$,
\[
\phi ^{(n,n+1)}\circ \varphi ^{(n)}=\kappa _{n}^{+}.
\]
$\varphi ^{(n)}$ also obeys
\[
(\Delta _{n}^{+}\otimes \kappa _{n+1}^{+})\circ \varphi ^{(n)}=(id_{n}\circ
\varphi ^{(n+1)})\circ \varphi ^{(n)}.
\]

Similarly we also introduce a family of algebras $\{\tilde{\mathcal{F}}%
^{(n)},n\in \mathbb{Z}\}$ in which $\tilde{\mathcal{F}}^{(i)}$ is given by
the relation
\begin{equation}
Y_{1}^{(i)}(v,\lambda )Y_{2}^{(i)}(u,\lambda )R^{(i+1)}(v-u,\lambda
)=Y_{2}^{(i)}(u,\lambda )Y_{1}^{(i)}(v,\lambda ),  \label{RYY}
\end{equation}
where $Y^{(i)}(u,\lambda )=(Y^{(i)}(u,\lambda )_{a})$. The analogue of $%
\varphi ^{(n)}$ is now denoted $\tilde{\varphi}^{(n)}$,
\begin{eqnarray*}
\tilde{\varphi}^{(n)}:\tilde{\mathcal{F}}^{(n)} &\rightarrow &\tilde{%
\mathcal{F}}^{(n-1)}\otimes \mathcal{A}_{n}, \\
Y^{(n)}(u,\lambda ) &\mapsto &Y^{(n-1)}(u,\lambda )\dot{\otimes}%
L^{(n)}(u,\lambda ).
\end{eqnarray*}
We require that $\tilde{\varphi}^{(n)}$ obey
\begin{eqnarray*}
\tilde{\phi}^{(n-1,n)}\circ \tilde{\varphi}^{(n)} &=&\kappa _{n}^{-}, \\
(\kappa _{n-1}^{-}\otimes \Delta _{n}^{-})\circ \tilde{\varphi}^{(n)} &=&(%
\tilde{\varphi}^{(n-1)}\circ id_{n})\circ \tilde{\varphi}^{(n)},
\end{eqnarray*}
where $\tilde{\phi}^{(n-1,n)}$ is the induced algebra homomorphism $\tilde{%
\phi}^{(n-1,n)}:\tilde{\mathcal{F}}^{(n-1)}\otimes \mathcal{A}%
_{n}\rightarrow \tilde{\mathcal{F}}^{(n-1)}$, and
\begin{eqnarray*}
\widetilde{\kappa }_{n}^{\pm }:\widetilde{\mathcal{F}}^{(n)} &\rightarrow &%
\widetilde{\mathcal{F}}^{(n\pm 1)}, \\
Y^{(n)}(u,\lambda ) &\mapsto &Y^{(n+1)}(u,\lambda ), \\
R^{(n+1)}(u-v,\lambda ) &\mapsto &\rho _{n+1}^{\pm }R^{(n+1)}(u-v,\lambda )
\end{eqnarray*}
is the lift of  $\rho _{n+1}^{\pm }$ onto $\widetilde{\mathcal{F}}^{(n)}$.
The operation $\tilde{\varphi}^{(n)}$ may be referred to as the \emph{dual
comorphism }with regard to the terminology comorphism used for $\varphi
^{(i)}$.

There is a natural pairing $\langle ,\rangle :\tilde{\mathcal{F}}%
^{(n)}\otimes \mathcal{F}^{(n+1)}\rightarrow End(\mathbb{C})$ between
elements of the algebras $\tilde{\mathcal{F}}^{(n)}$ and $\mathcal{F}%
^{(n+1)} $ given by
\[
\langle Y^{(n)}(u,\lambda ),X^{(n+1)}(u,\lambda )\rangle
=\sum_{a}Y_{a}^{(n)}(u,\lambda )X^{(n+1)a}(u,\lambda ).
\]
A crucial observation is that the operators
\[
\mathcal{T}^{(n)}(u,\lambda )\equiv \langle Y^{(n)}(u,\lambda
),X^{(n+1)}(u,\lambda )\rangle ,
\]
commute among themselves, {\em i.e.\ }
\begin{equation}
\lbrack \mathcal{T}^{(n)}(u,\lambda ),\mathcal{T}^{(n)}(v,\lambda )\rbrack
=0.  \label{TT}
\end{equation}
This equation can easily be obtained by multiplying together the
defining relations for $%
\tilde{\mathcal{F}}^{(n)}$ and $\mathcal{F}^{(n+1)}$ ({\em i.e.\
}equations (\ref {RYY}) and (\ref{RXX}) with upper indices
adapted), inserting the unitarity condition for the $R$-matrix,
and taking the pairing $\langle ,\rangle $ in both spaces labeled
by suffices $1$ and $2$.

Note that the commutation relation (\ref{TT}) will not be spoiled by the
action of $\varphi ^{(n+1)}$($\tilde{\varphi}^{(n)}$), due to the existence
of the algebra homomorphisms $\phi ^{(n,n+1)}$ and $\tilde{\phi}^{(n-1,n)}$
and the fact that $\kappa _{n+1}^{-}\circ \kappa _{n}^{+}=\kappa
_{n-1}^{+}\circ \kappa _{n}^{-}=id_{n}$. Therefore, we can act on $\mathcal{T%
}^{(n)}(u,\lambda )$ successively with the operators $\varphi
^{(n+1)}$, $\varphi ^{(n+2)}$, ..., $\varphi ^{(m-1)}$ as follows:
\begin{eqnarray*}
\mathcal{T}^{(n,m)}(u,\lambda ) &\equiv &(id_{n+1}\circ ...\circ
id_{m-2}\circ \varphi ^{(m-1)})\circ (id_{n+1}\circ ...\circ id_{m-3}\circ
\varphi ^{(m-2)})\circ ... \\
&&\circ (id_{n+1}\circ \varphi ^{(n+2)})\circ \varphi ^{(n+1)}\lbrack
\mathcal{T}^{(n)}(u,\lambda )\rbrack ,
\end{eqnarray*}
which results in operators $\mathcal{T}^{(n,m)}(u,\lambda )$ of the form
\begin{equation}
\mathcal{T}^{(n,m)}(u,\lambda )=\langle Y^{(n)}(u,\lambda
),L^{(n+1)}(u,\lambda )\dot{\otimes}...\dot{\otimes}L^{(m-1)}(u,\lambda )%
\dot{\otimes}X^{(m)}(u,\lambda )\rangle ,  \label{transfer}
\end{equation}
where $m>n\in \mathbb{Z}$. The conclusion is that the operators $\mathcal{T}%
^{(n,m)}(u,\lambda )$ still commute among themselves:
\[
\lbrack \mathcal{T}^{(n,m)}(u,\lambda ),\mathcal{T}^{(n,m)}(v,\lambda
)\rbrack =0.
\]

Since these operators carry a spectral parameter dependence, we
can expand them with respect to this parameter to get infinitely
many commuting operators. The operators
$\mathcal{T}^{(n,m)}(u,\lambda )$ may be viewed as generalizations
of the transfer matrix appearing in the usual quantum inverse
scattering method and thus are expected to yield novel
integrable/solvable models when the $L$-matrices and the $Y^{(n)}$ and $%
X^{(m)}$ are all specified in a specific representation.

\vspace{1pt}\textit{Remark}. In the case of standard (quasi-)Hopf
algebras, the structure which is analogous to the result of the
present section, known as the {\em comodule}, already exists
\cite{SK2}, and played an important role in constructing
integrable models with open boundaries. We emphasize that the
notion of comodule is only meaningful for algebras which are
co-closed. For our infinite Hopf family of algebras, co-closure
necessarily fails (otherwise we are back to the conventional Hopf
or quasi-Hopf algebra frameworks). So the comorphism is the best
structure we can design in mimicking the structure of the
comodule.

\section{Examples}

So far our study of the generalized $RLL$ algebras has remained on
the abstract level: we have not specified any concrete examples of
the operators $\rho _{n}^{\pm }$ and hence the algebra morphisms
$\tau ^{(n,m)}$. In this section, we provide such examples, which
may aid understanding of our earlier abstract constructions.

\vspace{1pt}For any input $R$-matrix $R^{(0)}$ there is of course
the trivial example given by $\rho _{n}^{\pm }=id$ for all $n$.
Actually $\rho _{n}^{\pm }=id$ implies $\tau ^{(n,m)}=id$ for all
$n,m$ and hence the corresponding co-structure degenerates into
the standard Hopf algebra structure. The following examples go
beyond these trivial cases. In particular, we are interested in
the cases which cannot be formulated as standard Hopf or
quasi-Hopf algebras or for which the formulations as standard Hopf
or quasi-Hopf algebras (if possible) are unknown.

First, let us discuss the possible actions of the operators $\rho
_{n}^{\pm }$. Since these operators preserve unitarity and the
Yang-Baxter equation, they cannot be chosen arbitrarily. However,
since the $R$-matrices have several arguments, including the phase
relative to the spectral parameter $u$, the deformation parameter
$\hslash $ and/or $\eta $ (which, if present, plays the role of
the trigonometric or elliptic period(s)) and the dynamical
variable $\lambda $, and so on, the actions of $\rho _{n}^{\pm }$
may be chosen in such a way that they change these in a consistent
way. For instance, the operators $\rho _{n}^{\pm }$ may

\begin{itemize}
\item  replace the trigonometric or elliptic period(s) by some other values $%
\eta ^{(n\pm 1)}$ ;

\item  shift the phase parameter by some amount $\pm \xi _{n}$ ;

\item  change the dynamical variable $\lambda $ to some other values $%
\lambda ^{(n\pm 1)}$, \textit{etc}.
\end{itemize}

In the following, we will not consider any dynamical $R$-matrices,
and hence no examples of the last kind will occur. However, as we
shall see, there are very rich choices of operators $\rho
_{n}^{\pm }$ even for the first two kinds only.

\subsection{in which $\rho_{n}^{\pm }$ act on the
trigonometric or elliptic period(s)}

\label{1}

In all the known cases of infinite Hopf families of algebras
realized through Drinfeld currents, the structure functions are
trigonometric or elliptic functions and the algebra morphisms
$\tau _{n}^{\pm }$ act by changing the period (or one of the two
periods) of the structure functions. From the
point of view of the $RLL$ realization, these cases correspond to operators $%
\rho _{n}^{\pm }$ which change the trigonometric or elliptic
period(s) of the $R$-matrix entries. For instance, in the case of
the trigonometric algebras ${\mathcal{A}}_{\hbar ,\eta
}(\hat{g})$, the operators $\rho
_{n}^{\pm }$ act on $R^{(n)}(u)$ by changing the period $\eta ^{(n)}$ into $%
\eta ^{(n\pm 1)}$, where
\begin{equation}
\frac{1}{\eta ^{(n+1)}}-\frac{1}{\eta ^{(n)}}=\hslash c_{n}.  \label{etan}
\end{equation}
In the case of $c_{n}=0$ for all $n$, all the periods $\eta
^{(n)}$ become identical, and the operators $\rho _{n}^{\pm }$ act
as the identity. Correspondingly the structure of the infinite
Hopf family degenerates into that of the standard Hopf algebra.

One should notice, however, that the difference between $\eta ^{(n+1)}$ and $%
\eta ^{(n)}$ need not necessarily depend on $c_{n}$ in order to
give rise to an infinite Hopf family of algebras. The operators
$\rho _{n}^{\pm }$ could just replace $\eta ^{(n)}$ by some
arbitrarily chosen $\eta ^{(n\pm 1)}$, which are completely
independent of $\eta ^{(n)}$. The infinite Hopf family of algebras
thus given will not degenerate into the standard Hopf algebra for
any value of $c_{n}$. The only known example of this kind that has
been studied in the past is the second family of algebras studied
in \cite{IHF3}.

\vspace{1pt}Notice also that the infinite Hopf family of algebras
given by operators $\rho _{n}^{\pm }$ can only be defined for
trigonometric or elliptic $R$-matrices.

\subsection{in which $\rho _{n}^{\pm }$ act on the phase parameter}

\label{2}

This is a novel class of examples which have not yet been studied
in the literature. To give readers a flavor of what the
corresponding algebras may look like, we give here an explicit
example. Take the trigonometric $R$-matrix

\[
R^{(0)}(u,\hslash ^{(0)},\eta )=\left(
\begin{array}{llll}
1 &  &  &  \\
& \frac{\sinh \pi \eta u}{\sinh \pi \eta (u+\hbar ^{(0)})} & \frac{\sinh \pi
\eta \hbar ^{(0)}}{\sinh \pi \eta (u+\hbar ^{(0)})} &  \\
& \frac{\sinh \pi \eta \hbar ^{(0)}}{\sinh \pi \eta (u+\hbar ^{(0)})} &
\frac{\sinh \pi \eta u}{\sinh \pi \eta (u+\hbar ^{(0)})} &  \\
&  &  & 1
\end{array}
\right)
\]
as input. Let the operators $\rho _{n}^{\pm }$ act by changing $\hbar ^{(n)}$
into $\hbar ^{(n\pm 1)}$, where
\begin{equation}
\hbar ^{(n\pm 1)}\equiv \hbar ^{(n)}\pm \xi _{n}.  \label{hbarn}
\end{equation}
Then, using the result of section three we can show that the
commutation
relations for the Drinfeld currents $E(u)$ and $F(u)$ of the algebra $%
\mathcal{A}(R^{(i)},R^{(j)})$ take the form

\begin{eqnarray*}
E(u)E(v) &=&\frac{\sinh \pi \eta (u-v+\hbar ^{(j)})}{\sinh \pi \eta
(u-v-\hbar ^{(j)})}E(v)E(u), \\
F(u)F(v) &=&\frac{\sinh \pi \eta (u-v-\hbar ^{(i)})}{\sinh \pi \eta
(u-v+\hbar ^{(i)})}F(v)F(u).
\end{eqnarray*}
Such algebras also do not degenerate into the standard Hopf algebra $c=0$,
provided $\xi _{n}$ are not proportional to $c_{n}$.

Notice that this kind of infinite Hopf family of algebras can be
defined for any type of $R$-matrices, elliptic, trigonometric, or
rational.

\subsection{in which all but a few of $\rho _{n}^{\pm }$ act as the identity}

\label{3}

The examples given in the last two subsections have the common
property that all the operators $\rho _{n}^{\pm }$ act in the same
way (though with different parameters for different $n$). One can
also think of cases in which the actions of the operators $\rho
_{n}^{\pm }$ vary drastically, so that one cannot describe the
actions of these operators by simple recursion formulae like
(\ref{etan}) or (\ref{hbarn}).

As a particular example of this kind, suppose that only a few of $%
\rho _{n}^{\pm }$ are different from identity. To be more concrete, let $%
\rho _{i}^{+}$ $(i=n,n+1,...,m-1)$ and $\rho _{j}^{-}$
$(j=n+1,...,m-1,m)$ be given as in subsections \ref{1} or
{\ref{2}, and the other $\rho _{n}^{\pm }$ be trivial, so that the
co-structure of the infinite Hopf family differs from the standard
Hopf algebra structure for only a few members of the family, {\em i.e.\
}$\mathcal{A}_{n},$ $\mathcal{A}_{n+1},$ $...,$ $\mathcal{A}%
_{m-1}$. As an extreme case, if $m=n+1$ there is only one algebra
in the family which has a non-trivial Hopf family structure; all
the other algebras (which actually degenerate into two different
algebras) are standard Hopf algebras. }

\vspace{12pt}

\centerline{* \hspace{1cm}*\hspace{1cm}*}

The special cases described in the above three subsections are
provided only as illustrative examples for the rich algebraic
structures which can be incorporated into the framework of
infinite Hopf families of algebras. We are far from being able to
list the variations for the choices of $\rho _{n}^{\pm }$
exhaustively. In some sense, the word ``infinite'' might best be
understood to refer to the infinite variation in choice of the
$\rho _{n}^{\pm }$, rather than to the size of each algebra family
(which may actually be finite in certain cases, as described in
the examples in subsection \ref{3}).

{}From the point of view of physical applications, one might try to apply the
examples given above in the framework of section \ref{S3} to get
infinitely
many commuting operators. As mentioned earlier, the operators $\mathcal{T}%
^{(n,m)}(u,\lambda )$ can be regarded as analogues of the transfer
matrices in integrable/solvable models. Their logarithmic
derivatives in finite dimensional representations are thus
expected to give rise to Hamiltonians \vspace{1pt}of solvable spin
chains. From the examples of subsections \ref{1} and \ref{2}, we
expect that certain solvable spin chains with site-dependent
couplings would arise. In contrast, the examples of subsection
\ref{3} are expected to give rise to spin chains with local
impurities or dislocations. Of course, these remarks make sense
only if a
finite dimensional representation for the matrices $\mathcal{T}%
^{(n,m)}(u,\lambda )$ is available.

\section{Concluding remarks}

The structure and representation theory of infinite Hopf families
of (super-)algebras have been studied in a number of papers over
the last few years. Most of the known results are concerned with
representations at $c=1$ in the Drinfeld current realization. By
use of the co-structure, it can be seen that representations of
such algebras at any $c\in \mathbb{Z}_{+}$ should exist. However,
the Yang-Baxter type realization for the infinite Hopf family of
(super) algebras has remained unknown until now.

In this paper, we proposed a generalized $RLL$ formalism for
generalized quantum affine algebras which are members of an
infinite Hopf family of (super-)algebras. The construction shows
that there is a very rich structure hidden in the generalized
formalism --- most of the known quantum affine algebras can easily
be seen to be special cases of the present formalism, and new
examples of infinite Hopf family of algebras can also be obtained
by specifying a particular set of operators $\rho _{n}^{\pm }$.

As we have stressed in the text, the co-structure of an infinite
Hopf family provides the possibility of defining novel classes of
commuting operators, which are essential ingredients in the
theories of integrable fields and/or completely solvable lattice
statistical models.

Despite the progress made in this paper, we would like to point
out some important open problems. One problem is the connection
with universal $R$-matrices. For standard Hopf algebras and
quasi-Hopf algebras, the universal $R$ matrix, which lives in the
tensor product space of the corresponding quantum algebra,
provides the algebraic foundation of $RLL$ formalism. The
analogous construction in our case is still unknown. Another
related open problem is the existence of representations at $c=0$
(here we are concerned
with the algebras which do not degenerate into standard Hopf algebras at $%
c=0 $). For standard quantum affine algebras, such (evaluation)
representations play a very important role because, on the one
hand, these representations are exactly where the universal
$R$-matrix evaluates to give rise to the numerical $R$-matrices,
and on the other hand the transfer matrices of solvable lattice
statistical models also take values in such representations. With
regard to physical applications we expect that the values of the
operators $\mathcal{T}^{(n,m)}(u,\lambda )$ studied in section
\ref{S3} would indeed become the transfer matrices of certain
lattice statistical models. If this is true, our generalized $RLL$
algebras would be an ideal algebraic tool to study spin chains
with site-dependent couplings and lattice statistical models with
local impurities or dislocations. We leave the detailed study of
this problem to future work.

\textbf{Acknowledgement.} L. Zhao would like to thank the Department of
Mathematics, University of York and the Abdus Salam International Centre for
Theoretical Physics for hospitality. Financial support from the Royal
Society of London, the UK PPARC, the Abdus Salam ICTP and the National
Natural Science Foundation of China are also warmly acknowledged.

\end{document}